# *Critical grain sizes and generalized flow stress –grain size dependence*

S.A.Firstov, T.G.Rogul, O.A.Shut (shut.ipms@gmail.com)

I.Frantsevich Institute for Problems of Materials Science, National Academy of Sciences of Ukraine, Kyiv, Ukraine

*The generalized equation that describes the yield stress dependence upon the grain size in the wide range of grain sizes has been obtained. There are two critical grain sizes ($d_{cr1}$, $d_{cr2}$) that correspond to the changes in strengthening mechanism. The equation includes Hall - Petch relation in the range of $d > d_{cr1}$. For $d_{cr1} > d > d_{cr2}$, the power in the Hall - Petch relation varies from -1/2 to -1. In a nanometer (nm) range ($d_{cr2} > d$) there are possibilities of softening (in case of "weak" boundaries) as well as significant strengthening (in case of "strong" boundaries).*

There have been enough discussions in the literature about the existence of so-called inverse Hall – Petch (HP) relation [1-4 etc.], i.e. the decrease of yield strength is observed with the reduction of the grain size after reaching a certain grain size $d_{cr}$.

For the description of the deviation from the classical HP - relation many models are proposed, which can be combined into four groups:
1) dislocation - based models [5, 6];
2) two – phase (grain body and grain boundary) - based models (a so-called 'mixture' rule) [7, 8, 9];
3) models of grain boundary plasticity mechanisms (diffusion - based models (Coble creep), grain – boundary – shearing models) [10, 11, 12];
4) models of wide stacking faults (models of twinning mechanisms) [13];
5) models that take into account the competition of different strengthening mechanisms, because, as a rule, several mechanisms act simultaneously in real materials [3].

In accordance with those models it's possible to describe the transition from Hall-Petch strengthening to the "inverse" HP relation at the some critical grain size $d_{cr}$. Kumar K.S. [14] proposed a scheme (fig. 1a) with two critical grain sizes $d_{cr1}$ *=100 nm* and $d_{cr2}$ *=10 nm*, in which the slope of *σ(d)* dependence changes.

Nevertheless, HP relation (1) is fulfilled well in the region of large grain sizes in the overwhelming majority of cases. Along with that, it is pointed out in numerous articles [7, 15, 16] for smaller grains that the strengthening at the grain refinement can be better described by the equation (2).

$$\sigma_1(d) = \sigma_0 + k_y d^{-1/2} \qquad (1)$$
$$\sigma_2(d) = \sigma_0 + k_1 d^{-1}, \qquad (2)$$

where the parameter $\sigma_0$ characterizes the averaged resistance to dislocation motion over the grain body and the coefficients $k_y$ and $k_1$ characterize the difficulty of slip transfer through the grain boundary.

The Thompson`s scheme [15], in contrast to fig 1a in [14], postulated not the slight slackening of the dependence of the yield strength upon the grain size, but, on the contrary, the increasing dependence of the yield strength takes place due to the transition from the dependence (1) to the dependence (2) at $d_{cr1}$.

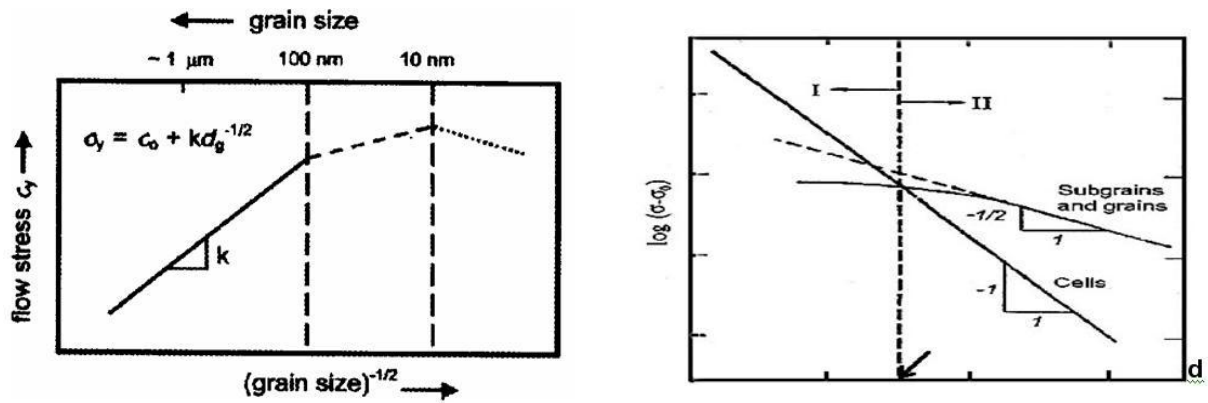

Fig.1 Dependence of flow stress on the grain size a) [14] , б) [15]

The simplest explanation of such transition is the following.
The coefficient $k_y$ is expressed as
$$k_y = \sigma_c r^{1/2}, \qquad (3)$$
where $\sigma_c = \alpha G b / L$ - the stress of the dislocation source actuation in the neighboring grain ($\alpha$ - coefficient of an order of 1, $G$ - shear modulus, $b$- Burgers vector, $L$ – characteristic length of a Frank – Reed source), $r$ is the distance of the source from the head of the dislocation pile-up. With the grain sizes in a range $d_{cr1} > d > d_{cr2}$ it is obvious that the length of dislocation source $L$ and its distance from the grain boundary $r$ can not be larger then $d$. Supposing that in such case $L \sim d$ and $r \sim d$, it can be shown that the coefficient is getting dependent upon the grain size as
$$k_y = k_1 d^{-1/2} \qquad (4)$$

The substitution of (4) into the equation (1) leads to the equation (2).

With the further grain size refinement ($d_{cr2} > d$), there might be the situation when the dislocation source can not generate dislocations at the stress lower than the theoretical shear strength. In this case flow stress dependence can be described as some expression $\sigma_3(d)$, which is determined by one of the concrete deformation mechanisms listed above. It is necessary to underline, that only for the dislocation - based models [5, 6], Coble creep [10, 11], and also for the so-called two – phase – based models [7, 8, 9] the analytic form for such dependences existed already.

Dislocation - based models can be excluded from the examination because the moving of the dislocations inside nanograins is quite difficult. The realization of the Coble creep at the room temperature for the nanocrystalline materials even for the copper needs anomalous high diffusion coefficient [12], which is unlikely for the metal with higher melting temperature.

From this point of view, for choosing an appropriate analytical dependence for $\sigma_3(d)$ (just like in [16]), we will select the approach based on a mixture rule:

$$\sigma_3(d) = \left(1 - \left(\frac{d-t}{d}\right)^2\right)\sigma_B + \left(\frac{d-t}{d}\right)^2 \sigma_V \tag{5}$$

Here $t$ – the thickness of grain boundary; $\sigma_B$ – the strength of grain boundary (GB –strength); $\sigma_V$ – the strength of the nanograin volume. Since the strength of the dislocation-free nanograin goes up to the limit (theoretical) strength, we can suggest that $\sigma_V$ is close to the theoretical strength ($\approx E/30$).

Earlier, Takeuchi used quite different formula for the yield stress [9]:

$$\sigma_y = \sigma_c - (\sigma_c - \sigma_B)\frac{\alpha t}{d},$$

$\alpha t \approx 2 nm$ - the effective thickness of the grain boundary. \hfill (6)

In our opinion, the expression (5) describes the composite behavior better, than the similar formula of Takeuchi (6). Besides, Takeuchi examined only the possible weakening of the dependence because the strength of the grain volume $\sigma_c$ is taken higher that the GB-strength $\sigma_B$, and he did not examine the variant, when the strength of the intergranular material can be equal to the strength of the grain body or even exceed it.

To receive the generalized dependence $\sigma(d)$ in the wide range of the grain sizes, the method that was used in [17] can be employed, where the distribution of the grains by size can be considered. At that, it is needed to consider the availability of two critical grain sizes, separating the manifestations of the strength mechanisms, described with the equations (1), (2) and (5).

Just like in the [17, 18], we will consider that the grain size log-normal distribution exists in the polycrystalline material:

$$f(V) = \frac{1}{V\sqrt{2\pi s_{\ln V}^2}} \exp\left[-\frac{(\ln(V) - m_{\ln V})^2}{2 s_{\ln V}^2}\right], \tag{7}$$

where $s_{\ln V}$, $m_{\ln V}$ - fitting factors.

Considering the existence of two critical grain sizes, the resulting yield stress of the polycrystal can be expressed as follows:

$$\sigma(d) = \frac{1}{m_v} \int_{v1}^{\infty} \sigma_1 V f(V) dV + \frac{1}{m_v} \int_{v2}^{v1} \sigma_2 V f(V) dV + \frac{1}{m_v} \int_{0}^{v2} \sigma_3 V f(V) dV \qquad (8)$$

where $V_1 = d_{cr1}^{3}$, $V_2 = d_{cr2}^{3}$

The plot corresponding to the equation (8) is built in the fig.2 for 4 different values of GB-strength $\sigma_B$ using chromium [19] as an example. Following data were used: the strength of the grain volume is closed to the theoretical value $\sigma_v = 12 GPa$; $\sigma_0 = 0.22 GPa$; $k_y = 1.6*10^{-2} GPa*mm^{1/2}$; the thickness of boundaries was $t = 2nm$; $d_{cr1} = 310 nm$ and $d_{cr2} = 67 nm$, $s_{\ln V} = 1$.

One can see, that three ranges of flow stress dependence vs the grain size can be distinguished. The first range extends to $d_{cr1}$, where the classical Hall-Petch`s equation is satisfied. The second one is up to $d_{cr2}$, where the sharp increase of strength vs the grain size follows $\sim d^{-1}$. In the nanometer range of the grain size, the decrease of $\sigma(d)$ with refinement of the grain size as well as it`s increase take place depending upon the change in GB-strength (from *E/100* to *E/15*, fig. 2).

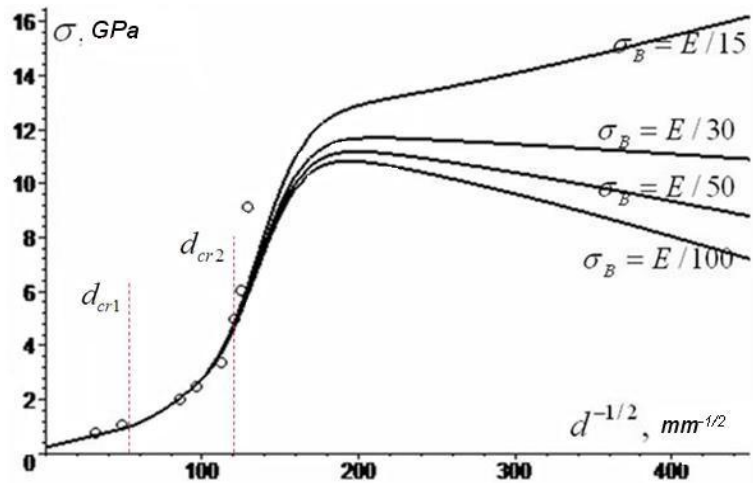

Fig.2 Flow stress vs. the grain size for some values of a GB- strength.
Experimental data taken from [19]

In the fig.3 there are experimental values *H/E* for the chromium-based materials with different grain sizes, which are produced with different technologies. The data for chromium films with the thickness of 40μm that are produced by magnetron cyclic sputtering are marked with green spots [19]. The red spots correspond to the chromium coatings with the thickness of about 6μm, that are produced by magnetron sputtering in the uniform regime and with the controlled adding of the oxygen during the sputtering (it was shown that oxygen segregates at the boundaries of grains) [21]. The blue spots correspond to the alloy Cr - 37.2 wt.% Ni with different thermal treatment [22]. It is obvious, that the grain boundaries will be different for all materials.

The dependencies shown in fig. 3 have been built according to the equation (8) considering the Marsh formula [20] for two values $k_y$ (max and min) and for two values of the GB- strength (for the case of relatively weak and relatively strong boundaries). The lowest dependence corresponds to the lower values $k_y = 1.4*10^{-2} GPa*mm^{1/2}$ for the chromium of high purity. The upper dependence corresponds to the higher value $k_y = 2.8*10^{-2} GPa*mm^{1/2}$.

According to the equation (8), it is supposed that at the change of the size of the grain, the structure and properties of the boundaries and, correspondingly $k_y$, do not change. Obviously, though, the state of the grain boundaries changes continuously for such material during grain size variations with thermal treatment. That is why, the experimental data shown in fig. 3, are located between two dependencies that reflect different values of $k_y$ for each grain size.

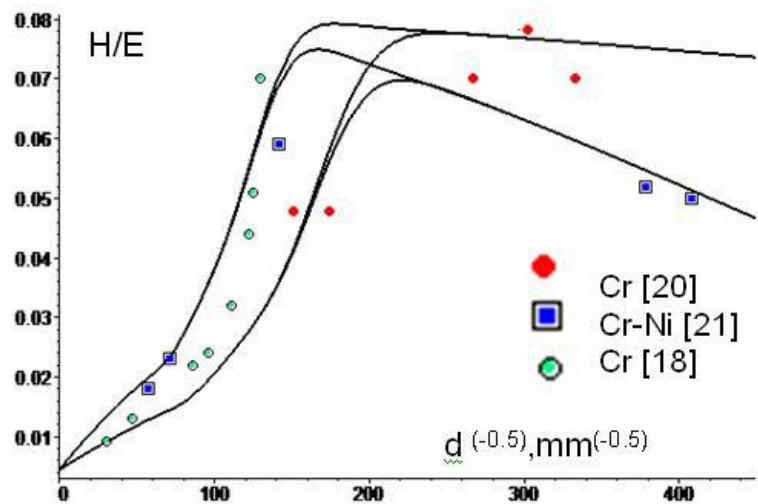

Fig.3. The generalized dependence of the normalized hardness (*H/E*) upon the grain size, which takes into account the possible changes of the boundaries state for two different $k_y$ and for two different GB-strength – $E/100 \div E/30$.

**Conclusion**

Summarizing, let us emphasize that, in general, there might be more than two critical sizes because strengthening mechanisms mentioned above might compete in the nano-region. Besides, not only the possible change in value of $k_y$ should be taken into account in certain cases but the change of the parameter $\sigma_0$ at the change of the grain size also.